\documentclass[twocolumn]{article}
\usepackage{fullpage}
\usepackage{amsmath}
\usepackage{graphicx}
\usepackage{hyperref}
\usepackage{verbatim}
\usepackage{pifont}
\usepackage{authblk}

\begin{document}

\title{Evolution as Explanation:  \\The Origins of Neural Codes and their Efficiencies}
\author[1,\ding{41}]{Han Kim} 
\affil{ Division of Biology and Biological Engineering, California Institute of Technology, USA}
\affil[\ding{41}]{ \texttt{hankim@caltech.edu}}
\date{\today}

\twocolumn[
  \begin{@twocolumnfalse}
    \maketitle
    \begin{abstract}
      \noindent Neural codes appear efficient. Naturally, neuroscientists contend that an efficient process is responsible for generating efficient codes. They argue that natural selection is the efficient process that generates those codes. Although natural selection is an adaptive process, evolution itself, is not. Evolution consists of not only natural selection, but also neutral stochastic forces that can generate biological inefficiencies. The explanatory power of natural selection cannot be appealed to, without regards for the remaining evolutionary forces. In this paper, we aim to reformulate the explanatory role of evolutionary forces on neural coding, with special attention to neutral forces. We propose a framework that argues for differing contributions of adaptive and stochastic evolutionary forces, for different phenotypic `levels', including those of neural codes. We assert that this framework is of special interest to neuroscience, because the field has derived much progress from an efficiency-based worldview. We advocate for a pluralistic neuroscience capable of appealing to both adaptive and non-adaptive explanations. 
      \\\\\\\
    \end{abstract}
  \end{@twocolumnfalse}
]

\section{Introduction}
Neuroscience seeks pluralistic explanations \cite{mayr_1961, tinbergen_1963, Krakauer:2017ie, hladky_2013}. In doing so, the field borrows from the ethos of multiple disciplines. From engineering-oriented fields, neuroscientists have learned to treat neural systems as reverse-engineering problems. Such approaches hypothesize overarching cost functions or goals, for neural systems, and test those hypotheses against neurobiological observations \cite{marr2010vision, Barlow_1961}.  Although an optimization-based worldview can be useful, biological features do not possess inherent goals or purposes \cite{Gould_Lewontin_1979, mayr_1997, dobzhansky_1973, Lynch_2007}. The tenets of evolutionary biology are consistent with this agnosticism, and accordingly, are incommensurable with an optimization-based worldview. For this reason, neuroscience's attempts to incorporate both the ethos of evolutionary-biological thinking and those of engineering disciplines have not been straightforward. This paper examines the nature of this pluralism in neuroscience. We argue that the current pluralism is ill-fitting, and reduces the evolutionary process to nothing more than some teleological rendition of natural selection. In response, we attempt to restructure the way in which neuroscientists invoke both optimization-based and historical process-based explanations, in the face of empirical results.   

In the current pluralism, neuroscience appeals to only a single evolutionary force---natural selection---to explain neurobiological phenomena. Namely, neuroscientists interpret natural selection as the process that generates proposed neurobiological efficiencies, such as neural codes \cite{Barlow_1961, barlow_2001, simoncelli_olshausen_2001}. In appealing to natural selection alone, however, neuroscientists not only curtail the full explanatory power of evolutionary-biological thinking, but at best, recast evolution as a force limited to natural selection. In actuality, evolution is not just natural selection, but consists also of neutral stochastic forces. We argue that these neutral forces can, in some cases, provide more useful explanations of neurobiological phenomena than optimization frameworks. 

On this note, we must mention, at the outset, an important corollary. This paper maintains that adaptive goal-oriented explanations are extremely useful. The hypothesizing of efficient codes yields much progress for neuroscience. Indeed, an automated appeal to adaptive explanations can be a sufficient heuristic, when thinking about the origins of neurobiological features. Like all types of explanations, however, optimization-based explanations have epistemological shortcomings. Namely, the optimization thesis is unfalsifiable, because we can interpret all observable phenomena in optimization terms. Either we have already derived a fitting cost function that the observations appear to match, or if not, we can contend that there still exists some appropriate function that has not yet occurred to us. Our position is that when faced with the latter case, the more useful approach---or at least an approach that merits serious consideration---is to interpret the observed phenomenon as the result of a historical process agnostic to any speculative goal or function. We contend that neutral evolutionary forces are one such explanatory process. Importantly, we acknowledge that historical non-adaptive explanations are also unfalsifiable. All empirical observations can be interpreted as the result of some preceding events from its history. In other words, this paper is not about denigrating one type of explanation, in favour of another---both have merits. Rather, we are concerned with deriving appropriate contexts for appealing to each explanatory class. 

We begin the paper by examining the history of the efficient coding hypothesis. We argue that the hypothesis tacitly comprises of two types of conflated explanations: a goal-based explanation, typical of engineering disciplines, and a historical process-based explanation, typical of evolutionary biology. We contend that making this distinction explicit is worthwhile. It reveals how neuroscientists currently view the explanatory capabilities of evolution. We then articulate the explanatory usefulness of non-adaptive evolutionary forces. We show that under certain conditions---namely, when species possess small effective population sizes---neutral stochastic forces dominate a species' evolutionary trajectory. We provide established examples from molecular biology and genomics that convey the explanatory power of non-adaptive forces. We then survey the contrasting efficiencies observed in neural coding. Given the overwhelming evidence that viewing neural codes in terms of efficiency is fruitful, we ask under which circumstances we should incorporate non-adaptive explanations. We articulate one evolutionary biological framework that considers both adaptive and non-adaptive explanations. We argue that the appeal to either adaptive or non-adaptive explanations should depend on both the phenotypic `level' that one is interested in, as well as the species' effective population size. In doing so, this framework is consistent with both the efficient codes argued in systems neuroscience, as well as the non-adaptiveness conveyed in genetic codes. We note that this framework can be applied to the thinking about phenotypic evolution across the general biologies. We contend, however, that it is of special interest to neuroscience, because unlike other biological disciplines, an efficiency-based worldview has provided neuroscience with immense progress. We provide neurobiological examples consistent with the proposed framework. We close in support of a pluralistic neuroscience that accommodates for both adaptive and non-adaptive explanations. 

\section{Two types of explanations and the efficient coding hypothesis}

\begin{figure*}[htp]
\centering
\includegraphics[width=15 cm]{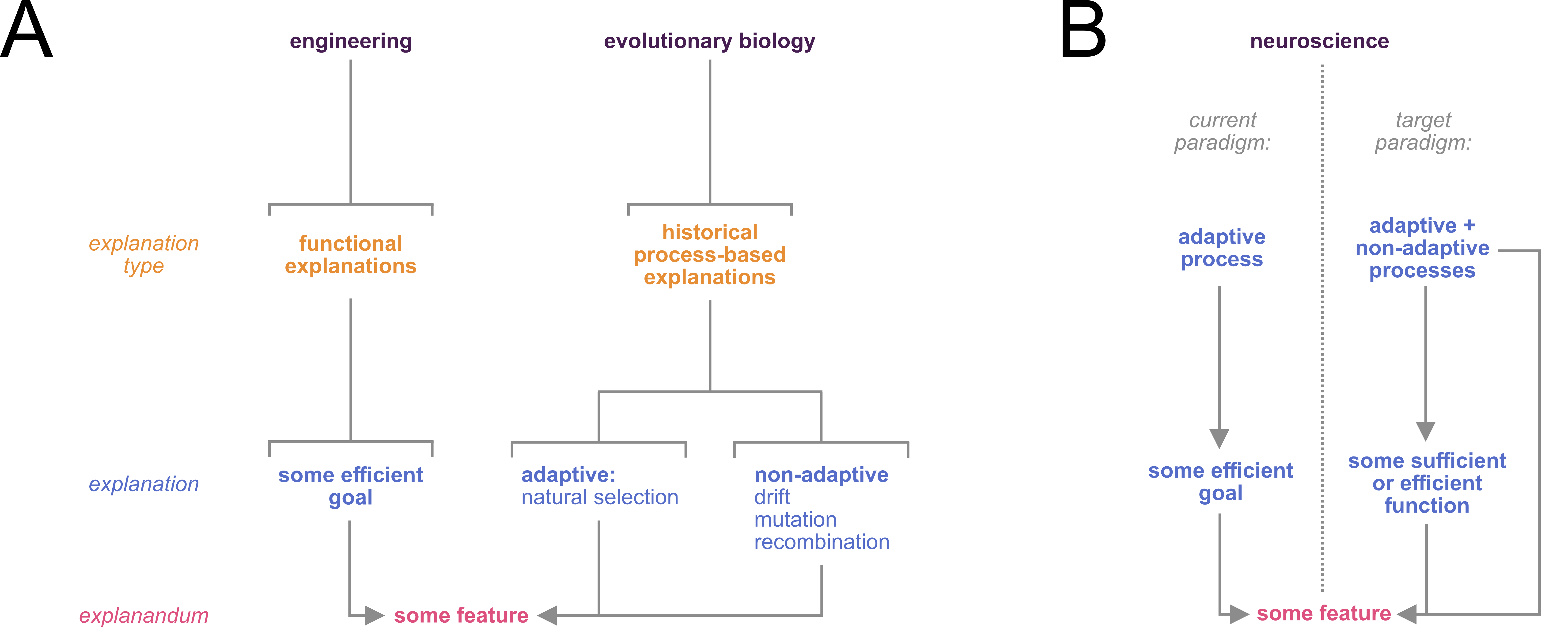}
\caption{\textbf{Different disciplines tend to use different types of explanations to explain observable phenomena.} A) In engineering, functional explanations that reference some efficient goal are used to explain some observed feature. In evolutionary biology, historical processes, which are either adaptive or non-adaptive, are used to explain some observed feature. B) In neuroscience, both adaptive and historical process-based explanations are used. Currently, however, we argue that adaptive processes are used to explain the existence of some hypothesized cost function, which is used to explain some observed feature. We argue that a more useful epistemological framework is to refer to both adaptive and non-adaptive processes, as well as functional explanations, to explain some observed feature. Vertical braces denote the bottom term being a subset of the top term. Terms before arrows explain terms after the arrow.}
\label{fig_explanations}
\end{figure*}

Thinkers of the natural world contend that multiple types of explanations can explain a given observation \cite{2017physics, aristotle1998metaphysics, mayr_1961, tinbergen_1963, hladky_2013}. Although there exists a rich history on the divvying of explanatory types, in this paper, we will categorize explanations of the biological world in one of two classes (Figure \ref{fig_explanations}A). The first is functional. Biological features are the way they are because they help achieve some supposed cost function or goal. This type of explanation is typical of engineering disciplines and can be immensely useful. They are popular with systems neuroscientists. The second type of explanation is historical. Biological features are the way they are because of preceding events in their generative process. This type of explanation is traditional to developmental biologists and evolutionary biologists, and can also be immensely useful. It has also found its way into systems neuroscience. Recognizing this explicit distinction between explanation types is necessary for understanding how neuroscientists use evolution to explain phenomena. 

An examination of the efficient coding hypothesis reveals how neuroscience employs both function-based and historical process-based explanations. In Barlow's original articulation of the hypothesis, he asks ``what are sensory relays for?" \cite{Barlow_1961} Since then, the hypothesis has successfully explained many empirical observations, and has spread to other aspects of neural coding. It can be found in relation to not only early sensory processing, but also for more central neural processes \cite{olshausen_field_1996, Pitkow_Meister_2012, brunton_2013}, as well as motor efferents \cite{hoyt_taylor_1981, sparrow_newell_1998, Mimica_Dunn_Tombaz_Bojja_Whitlock_2018}. In all these cases, the spirit of the original hypothesis remains intact: what is the neural code \textit{for}? In other words, the hypothesis squarely concerns itself with devising a function-based explanation, and in doing so, assumes an implicit goal for the nervous system's properties. We emphasize that there is nothing wrong with this position. It yields much explanatory insight for many neurobiological phenomena. 

The logic behind the efficient coding hypothesis, however, is not limited to a function-based explanation. It also tries to explain \textit{how} an organism might end up with an efficient code of sorts (Figure \ref{fig_explanations}B). To do so, it first answers its original question, ``what are sensory relays for?" It claims they are for representing the animal's natural environment. It hypothesizes that the animal likely does so `efficiently'---perhaps by using minimal neural impulses---such that the animal best represents its natural environmental statistics, over some non-ecological statistics. 

Given that efficient codes exist to optimally represent the animal's natural environment, the efficient coding hypothesis then returns to ask how such a code could arise. It seeks to link the natural environment to the qualities of the organism. To do so, it very intuitively invokes the process of natural selection. The efficient coding hypothesis argues that, over historical time, the process of natural selection alters the structure of a species' neural computations such that it efficiently represents its native environment. In other words, the hypothesis invokes a historical process to explain how an organism ends up with an efficient code. It explains neurobiological observations via a function-based explanation, and it employs a process-based explanation to explain how exactly the code meets its function. 

This logic, however, risks a nuanced but meaningful misconception. Natural selection, although an adaptive process, is explicitly not goal-oriented. In the words of evolutionary biologist, Ernst Mayr, natural selection and other ``[h]istorical processes ... can \textit{not} act purposefully" \cite{mayr_1961}. The efficient coding hypothesis' invoking of a non-goal-oriented process, in order to explain how something achieves its supposed goal, risks recasting the natural selection process in teleological terms. Indeed, we think that as a result of the coding hypothesis' articulation, many neuroscientists tacitly interpret natural selection as a goal-oriented process. Although we reject the notion that natural selection could be at all purpose-driven, we do not disagree with the idea that natural selection contributed to the rise of purported efficient codes in the peripheral nervous systems. Natural selection, as it was originally conceived---as an undirected and iterative process that imposes ecological constraints on a species---perhaps provides the best explanation for many efficient neural codes. We simply wish to articulate the necessary evolutionary biological caveat that has been amiss from the incarnations of the efficient coding hypothesis: although natural selection might explain efficient codes, natural selection is not an efficient process with respect to some goal. Goals and adaptive processes are two different things.

We contend that the overlooking of this subtlety persists in the many articulations and applications of the efficient coding hypothesis, and is the culprit behind neuroscience's limited means of appealing to evolution as an explanation. In the worst of such cases, neuroscientists invoke natural selection as an epistemological panacea. They use it to explain how a neural system wound up meeting whatever cost function objective they are interested in, even when the code itself does not appear to be particularly efficient. For example, several decades after Barlow published the efficient coding hypothesis, he revisits his formulation, and argues that he was wrong in over-emphasizing the role of `compression' in an efficient code, and contends that redundancy in neural coding is a crucial function for efficient nervous systems \cite{barlow_2001}. Although we agree with this reasonable amendment, we disagree with Barlow's continued, and if anything, more explicit appeal to natural selection as something that explains a redundant neural code. The continued allegiance to natural selection in Barlow's revisit may be a vestige from the fact that his original efficient coding hypothesis invoked an adaptive evolutionary process, albeit under more sensible circumstances. Alternatively, or perhaps additionally, Barlow and other neuroscientists are unaware of alternative evolutionary forces that may better explain their phenomena. In the next section, we introduce such alternative forces. We articulate how evolution is more than just natural selection, and we describe how non-adaptive neutral forces can sometimes explain phenomena more usefully than an optimization-based framework.

\section{Evolutionary biology usefully explains inefficient codes}

Even in the face of biological features that we might intuitively interpret as inefficient---such as redundancy in neural representations---we can still appeal to goal-oriented explanations. For example, we can argue that the observed redundancy aids the system in achieving the goal of robustness, perhaps from noise or from external perturbations. More generally speaking, we can always formulate a goal-oriented explanation about an observation because we can claim that either the observed inefficiency is exactly the point of the system, or because we can argue that we do not yet know the goal of the system, but the system most certainly has a goal, and with further empirical evidence, such a goal will become clear. We contend that adopting these epistemological positions is limited. As an alternative, we introduce non-adaptive evolutionary forces, which we contend are more useful than efficiency-based frameworks for explaining inefficiencies. We begin by introducing the explanatory power of non-adaptive forces in its native context. The goal of this section is, in part, pedagogical. We aim to explain key evolutionary biology concepts, without being mired by technical derivations. 

\subsection{The effective population size defines drift}

Evolution is not natural selection. Rather, it consists of four forces, of which natural selection is just one. The other three---mutation\footnote{A novel change in genetic sequence. In nature, is the result of a random process.}, recombination\footnote{The rearrangement of genetic material. In nature, is the result of a random process.}, and genetic drift\footnote{The random sampling of genetic material from the parental generation, to produce the offspring generation. Results in differing genetic compositions between the parental and offspring generations.}---are non-adaptive and neutral because they are random and do not depend on an animal's fitness properties. Under what conditions, then, will neutral forces provide the most explanatory power regarding a biological feature's origins? We begin with drift. 

Sewall Wright, a forefather of population genetics, first sketched the concept of the effective population size ($N_e$) to provide a way of calculating the rate of evolutionary change caused by the random sampling of allele\footnote{One of two or more versions of a gene. They arise by mutation.} frequencies---that is, genetic drift \cite{wright_1931}. An intuition for $N_e$, and how it differs from the `regular' population size of breeding individuals ($N$), is essential for understanding neutral forces. $N_e$ is best understood in a highly idealized but important type of random sampling called the Wright-Fisher population \cite{wright_1931, fisher_1923, fisher_1931}. The Wright-Fisher population assumes a randomly mating population that consists of a number of diploid\footnote{A cell or organism with two complete sets of genetic information. If the cell or organism has only one complete set of genetic information, it is haploid.} hermpahroditic individuals. In the Wright-Fisher population, these individuals are the total number of breeding individuals, $N$. They reproduce with discrete generations, and each generation is counted at the time of breeding. The individuals of each new generation are the result of random sampling, with replacement, from the gametes of the parents, and the parents die immediately after reproduction. If we accept all the idealized assumptions of the Wright-Fisher population, then the rate at which drift, or random sampling, results in a differing genetic composition in the offspring is $\frac{1}{2N}$. The coefficient `2' comes from the fact that the Wright-Fisher population assumes that the parents are diploid, so each individual actually has two inheritable genetic sets. If the members of the Wright-Fisher population were haploid or triploid, then the rate of change from drift across the discrete generations would be $\frac{1}{N}$ or $\frac{1}{3N}$, respectively. In other words, the goal of the Fisher-Wright population is to identify a feasible set of biological assumptions that result in the most simplified expression for genetic drift. This theoretical concept provides us with an intuition for drift, and as we will see, for $N_e$. 

Most biological populations, however, do not reproduce in a fashion similar to the sampling from a bag of marbles. For example, many species have two sexes, select mates non-randomly, and will not produce the progeny of the next generation all at once. For these more complex situations, we invoke the concept of the \textit{effective} population size, $N_e$. The effective population size has, within it, all the relevant biological complications that distinguish this population from a Wright-Fisher population, such that we can simply substitute $N$ for $N_e$, for most problems we wish. For example, in a diploid non-Wright-Fisher population, we might say that the rate of change in genetic composition due to drift is $\frac{1}{2N_e}$ instead of $\frac{1}{2N}$. In other words, we defer. The consideration for any biological complications when working with a given population genetic expression originally meant for a Wright-Fisher population is offloaded from that particular expression and its use of $N$, to its substitution for $N_e$. Many theoretical and computational methods, such as coalescent theory, exist for inferring $N_e$ for a population \cite{charlesworth_2009}, but they are outside the scope of this paper. The reader needs know only that the effective population size, and more specifically, $\frac{1}{N_e}$, approximates the neutral force that is drift\footnote{Some authors distinguish between various types of effective population sizes, such as $N_e$, $N_l$, and $N_g$, which all take into account different factors such as timescales, genome linkage effects, and the ploidy of a species \cite{Lynch_2007, lynch2007origins}. For our purposes, we will not be making such distinctions, because they do not matter for our arguments with the reader. We will use exclusively $N_e$ when referring to the effective population size.}. 

\subsection{The selection coefficient expresses the strength of natural selection}

In order to understand when neutral forces outweigh adaptive forces in evolution, we need to introduce the selection coefficient, $s$. Doing so requires an understanding of fitness. In the broadest sense, fitness involves the ability of organisms or populations to survive and reproduce in their environment\cite{orr_2009}. The concept of fitness is important for natural selection and the selection coefficient, because natural selection requires differing levels of fitness across members of a population, and that some of those fitness differences be inheritable---that is, that they have a genetic basis. For this reason, we will express our understanding of fitness in more specific genetic terms. 

Evolutionary biologists distinguish between two types of fitness metrics\cite{orr_2009}. One is the absolute fitness, $W$. It refers to a genotype's expected total fitness, and encapsulates the complexities that come with integrating all imaginable biological properties, such as viability, mating success, and fecundity. $W$ must be greater than or equal to zero. The other type of fitness, which is more commonly used in evolutionary biology, is relative fitness, or $w$. The relative fitness of a genotype is simply its absolute fitness, but normalized in some way. Its most common normalization is dividing the absolute fitness of a genotype by the absolute fitness of the fittest genotype. In doing so, $w$ is often bound between 0 and 1, where 1 is the fitness of the fittest genotype. From the definition of $w$, we can easily understand the selection coefficient, $s$. It is simply a genotype's $w$, relative to the fittest genotype's $w$. That is, $s = 1 - w$, where $w$ is the relative fitness of some genotype. So if some genotype is incredibly fit and has $w=1$, then for that genotype, $s=0$, and selection against that genotype is non-existent. Conversely, if some genotype is extremely deleterious and has $w=0$, then for that genotype, $s=1$, and selection against that genotype is total. The genotype will contribute nothing to the subsequent generation. From these examples, we can see how the value of $s$ expresses the strength of natural selection. Like many terms in evolutionary biological theory---such as $N_e$---several methods exist for empirically estimating $s$\cite{orr_2009}, but they are outside the scope of this paper. 

\subsection{Random forces can dominate evolution} \label{Random forces can dominate evolution}
From our above definitions of drift and natural selection, we can formulate an expression for evolution that considers the contributions of both non-adaptive and adaptive forces. Given that we can articulate the strength of drift as $\frac{1}{N_e}$, and the strength of natural selection as $s$, we can define their relative contributions as $\frac{\text{selection}}{\text{drift}} = \frac{s}{\frac{1}{N_e}} = N_es$. 

To appreciate the $N_es$ ratio, we must consider a classic case of a diploid population species, such as \textit{Drosophila melanogaster} or \textit{Mus musculus}. In such a species, a given gene might have two possible alleles, \textbf{A} and \textbf{a}, where allele \textbf{A} might have a slight selective advantage, \textit{s}, over \textbf{a}. Given that one allele is advantageous over the other, a worthwhile question that directly relates to our original intents, is to ask under which scenario the advantageous version of the gene, \textbf{A}, will become predominant, or fixate, in the population. In other words, given the selective advantage that \textbf{A} has over \textbf{a}, will natural selection always ensure the fixation of \textbf{A}? The answer to this question is no. Let's say that the mutation rate of \textbf{a $\rightarrow$ A} is $m$ times greater than the mutation rate of \textbf{A $\rightarrow$ a}. If so, then standard population genetics theory\cite{kimura1983neutral, guy_hirsh_2005, Lynch_2007, lynch_hagner_2015, lynch_2020} argues that the probability of fixing a mutation to \textbf{A} is $e^{2N_es}$ times more likely than fixing a mutation to \textbf{a}. That is, the ratio of probabilities of being \textbf{A} versus \textbf{a} is $me^{2N_es}$. This relationship is particularly interesting, because when $\frac{1}{N_e} \gg |s|$, the probability of fixing \textbf{A} over \textbf{a} in the population approaches exactly the ratio of mutation rates, $m$. That is, when $\frac{1}{N_e} \gg |s|$, or said another way, when the effective population size, $N_e$, is extremely small, then the likelihood of fixation for some advantageous allele is a function of mutation pressures alone. Under small $N_e$, random forces dominate the course of a population's evolution. We can plot this relationship, and observe the effects of altering the $N_es$ ratio (Figure \ref{fig_neutralforces}A). Small $N_es$ ratios result in evolutionary trajectories that are nothing more than the product of mutation rates, a neutral stochastic force. 

\subsection{Empirical support for predictions on biological coding inefficiency} \label{Empirical support for predictions on biological coding inefficiency}

\begin{figure*}[htp]
\centering
\includegraphics[width=16.5 cm]{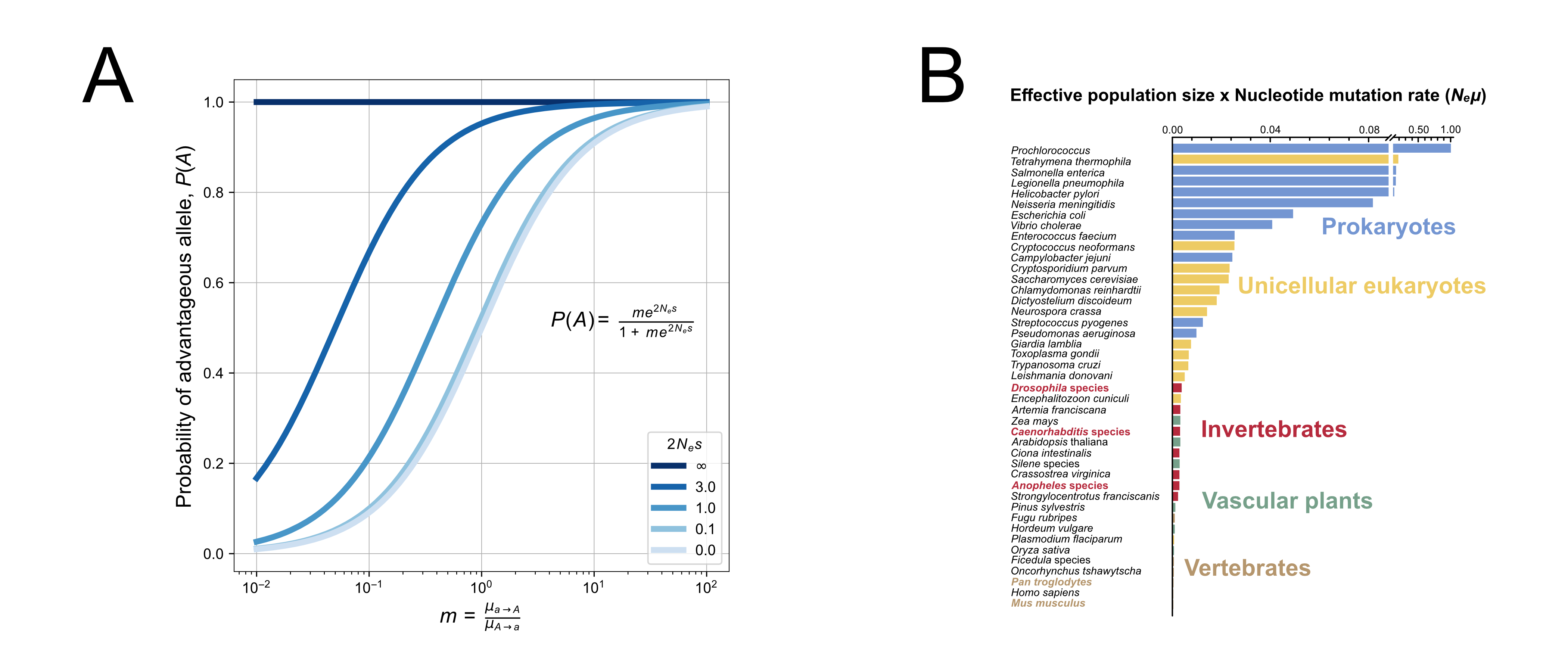}
\caption{\textbf{Stochastic forces make non-negligible contributions to evolution.} A) As the effective population size increases, relative to the selection coefficient, then the probability that an advantageous allele fixates will approach the fixation probability driven by mutation rates alone. This figure is a modified reproduction \cite{Lynch_2007}. B) Organisms with nervous systems possess smaller effective population sizes. Values of $N_e\mu$ for different species are depicted, from which $N_e$ values can be estimated, because $\mu$ ranges, at most, by an order of magnitude across known phyla (see text in Section \ref{Empirical support for predictions on biological coding inefficiency} for details). In bold and in colour on the x-axis are popular model organism species for neuroscience research. This figure is a modified reproduction \cite{Lynch_Conery_2003}.}
\label{fig_neutralforces}
\end{figure*}

From the above pedagogy, one might ask what sorts of populations or species possess small effective population sizes? Such species are likely more susceptible to neutral forces than species with larger $N_e$ values. To answer this question, we must introduce another parameter from evolutionary biology: $N_e \mu$. This composite parameter is the effective population size, $N_e$, scaled by the mutation rate, $\mu$. It determines the standing equilibrium level of approximately neutral genetic variability in a population \cite{kimura1983neutral, Lynch_Conery_2003, charlesworth_2009}. Biologists can estimate this value empirically, by quantifying the extent of synonymous mutations\footnote{Mutations in the genetic sequence that do not result in the encoding of different amino acids. For example, both GCA and GCG genetic triplets, or codons, encode for the alanine amino acid, even though the sequences differ by a base pair. The GCA $\rightarrow$ GCG mutation is synonymous or silent.} in a species, because synonymous mutations are neutral \cite{Lynch_Conery_2003}. Given that across species, the mutation rate per base per cell division, $\mu$, ranges from $5\times 10^{-11} \text{ to } 5\times10^{-10}$ \cite{drake_1998}, that is, by a single order of magnitude, we can factorize and interpret $N_e\mu$ as $N_e$. Doing so results in estimates of $N_e$ for prokaryotes that is at least $10^8$, for unicellular eukaryotes that is $10^7$ to $10^8$, for invertebrates that is $10^5$ to $10^6$, and for vertebrates that is $10^4$ to $10^5$ \cite{Lynch_Conery_2003}. A plot of $N_e\mu$ estimates for various species is shown in Figure \ref{fig_neutralforces}B. In other words, the effective population sizes of nervous system-possessing animals is far smaller than those without. They are more susceptible to random genetic drift than their neuron-less counterparts. Within nervous system-possessing animals, we see that invertebrates are less privy to random drift than vertebrate animals\footnote{An interesting observation for the invertebrate neuroscientist. Its consequences are slightly outside the scope of this paper.}. We also note that, in general, estimates of $N_e$ are much smaller than estimates of the true number of breeding individuals, $N$ \cite{lynch2007origins, crow_newton_1955, frankham_1995}. In other words, most biological populations do not merely differ from Wright-Fisher population dynamics, but do so in a way that promotes the effects of drift. 

We have provided both theoretical and empirical evidence for thinking that non-adaptive evolutionary forces are at least worth acknowledging, when thinking about nervous system evolution. In addition, however, observed biological features can also make more sense under a view of evolution that incorporates stochastic forces. There are several established examples in relation to genetic coding. Prokaryotes possess the largest $N_e$ values, and accordingly, their genome architectures are remarkably efficient. Nearly all of their genomes are dedicated to coding proteins and they possess virtually no non-coding sequences, such as introns and non-coding RNA \cite{lynch2007origins, milo2016cell}. In addition, genes that function in common biological processes are found adjacent to each other in the genome, such that the organism can readily achieve co-expression of those related genes \cite{overbeek_matlsev_1999}. Prokaryotic genes can also be polycistronic, meaning that a single mRNA transcript can encode multiple proteins at a time \cite{kozak_1999}. 

In contrast, genetic coding in multicellular eukaryotes appear to be anything but efficient. Eukaryotes are monocistronic, and so more energy and molecular machinery must be expended to transcribe and translate a given coding unit \cite{kozak_1999}. Moreover, the majority of multi-cellular eukaryotic genomes are non-coding elements. Less than 2\% of human genes are protein-coding, whereas 90\% to 100\% of prokaryotic genomes encode for functional units \cite{milo2016cell}. The dynamics of coding in multi-cellular organisms is also remarkably inefficient. Most of the eukaryotic genome is transcribed, and yet only a small fraction of those transcripts undergo maturation and subsequent translation into proteins \cite{menet_rosbash_2012}. Transcripts not processed for maturation and translation are digested and decayed \cite{mcnicoll_neugebauer_2014}. The large investment of energy in producing intermediary products from nearly all of the genome, only to have most of those intermediates degraded, illustrates a highly inefficient biological coding process. 

Evolutionary biology can explain these observations without necessarily invoking efficiency and natural selection \cite{Lynch_2007}. Seemingly inefficient biological qualities are nothing more than the outcomes contributed by random sampling processes. Populations with smaller $N_e$ are simply more prone to such processes. In contrast, when faced with putative inefficiencies, optimization-based frameworks such as those seen in engineering disciplines are resigned to one of the two positions mentioned at the very beginning of this section: either the observed inefficiencies are, for some reason, exactly the point of the system, or that the analyst has wrongly identified the goal of the system. We maintain that these two positions are epistemologically limited. They do not explain the observed phenomena by appealing to some scientific process, but rather, defer the explanation to some function that is not yet understood or known. We reassert, however, that optimization-based frameworks still have an essential place in biology and neuroscience. To demonstrate this point, we provide a brief review of the overwhelming evidence in favour of an optimization framework in neuroscience. 

\section{Efficient codes and nervous systems}

Given the relatively small effective population sizes seen in nervous system-possessing animals, we might expect abundant inefficiencies across neural codes. This hypothesis, however, appears to be exceptionally false. Optimization frameworks can clearly provide much explanatory power, especially in neuroscience. In this section, we provide the reader with diverse examples from across the nervous system that support the epistemological benefits of assuming purposes in natural systems. 

\subsection{Sensory systems}
In feline, human, and grasshopper auditory systems\cite{Machens_Gollisch_Kolesnikova_Herz_2005, smith_lewicki_2006}, sensory systems appear to use a sparse spike code to represent the acoustic structure of a given stimulus. Kernel functions for an optimal sparse representation learned from auditory stimuli closely approximate the physiological reverse-correlation filters \cite{smith_lewicki_2006}. Importantly, these efficient codes only succeed in using a sparse representation when the auditory stimulus is derived from natural scene statistics. In other words, efficient neural codes appear to at least be correlated with the germane particulars of the animal's natural environment. One reasonable interpretation to the exquisite match between an animal's native environment and the neural code that processes that environment is that the nervous system is somehow tuned or optimized for the environment. A simple process-based and folk-biological explanation of these findings might be that selective forces encompass ethological cues, and animals that fail to efficiently respond to those cues exhibit survival-inappropriate behaviours, and are selected against. Indeed, efficient codes tend to match not only the ethological sensory space, but are especially tuned for those stimuli that are behaviourally relevant\cite{Machens_Gollisch_Kolesnikova_Herz_2005, machens_herz_2001}. 

We observe theoretical and experimental evidence for efficient neural codes in modalities apart from audition. For example, a survey of ommatidia diameter and eye height from 27 \textit{Hymenoptera} species suggest that their facets have evolved to maximize visual sampling, while minimizing the blur from diffraction \cite{barlow_1952}. This conclusion derives from a fundamental physics principle showing that the optimal resolving power of a lens with diameter $d$ is equivalent to $\sqrt{a\lambda}$, where $a$ is the angle subtended by two point sources that can still be detected as double, and $\lambda$ is the wavelength of incoming light. The surveyed insects indeed possess a linear and proportional relationship between $\sqrt{a}$ and $d$ \cite{barlow_1952}. In terms of neural coding, the retinal ganglion cells of both salamanders and macaques decorrelate spatial features of visual inputs, as a means of achieving sparse spiking in the retina and optimizing visual coding efficiency \cite{Pitkow_Meister_2012}. 

\subsection{Motor systems}

We also find efficiency at the opposite end of the periphery, although not necessarily with respect to neural coding. Instead, most examples of efficient motor control are seen in terms of the amount of energy expended in muscles, to achieve a particular task. Given that the amount of muscle energy required for a behaviour vastly exceeds those needed for neural spiking \cite{ortega_2015, sengupta_2010}, one can argue that the relevant cost function at the motor periphery should relate to the joules of muscle work expended, rather than some sparse spike code. Classic work on horse gaits show that freely moving horses self-optimize their movement speeds. Under naturalistic conditions, they largely operate in locomotor regimes that require minimal amounts of consumed oxygen for moving some unit distance. These results hold across multiple gait types, such as walks, trots, and gallops \cite{hoyt_taylor_1981}. Similar results of self-optimization with respect to gross energy expenditure have also been reported in human ergometer studies \cite{sparrow_newell_1998}. 

Despite the foremost relevance of muscle expenditure over neural spiking for motoric efficiency, some recent evidence suggests that the encoding of naturalistic behaviours may also have some efficient basis. Simultaneous microdrive recordings and 3D posture tracking of the head, neck, and back of freely moving rats reveals that proportionately fewer cells in the posterior parietal and motor cortices fire when the animal partakes in common `default' postures \cite{Mimica_Dunn_Tombaz_Bojja_Whitlock_2018}. The encoding of naturalistic poses appears to use a minimal number of cells to represent the repertoire of ethological motor ensembles. 

\subsection{Central systems}

Further examples of efficiency have been argued in deeper processing regions, albeit often with different, and arguably more ambiguous, objectives from the sparse coding seen at the periphery \cite{simoncelli_2003}. Learning algorithms that maximize sparseness, for example, succeed in recapitulating the spatially localized, oriented, and bandpass features of mammalian visual cortical cells \cite{olshausen_field_1996}. In general, sparse codes in the visual cortex appear to be useful for learning and processing incoming spike patterns \cite{olshausen_field_1996}, or for parsing large amounts of signal from background noise \cite{ringach_malone_2007}, in the face of overcomplete representations. Even though these kinds of objective functions differ from those efficient representations seen in the periphery, where the goal is to minimize the number of spikes needed for expressing some maximal description of the environment, both scenarios still offer examples of efficient coding. 

Operating on the assumption that nervous systems have innate goals or functions is undeniably useful for neuroscience. An appeal to natural selection as the historical process that explains these findings is also not problematic. In fact, it is likely the most appropriate and useful type of process-based explanation for these observations. In light of the previous section's pedagogy, however, the reader may ask themselves two questions: first, why do small $N_e$ species, such as those with nervous systems, possess efficient neural codes, and second, whether there is a general place or context under which an appeal to neutral evolutionary processes is most appropriate. We directly address this question in the next section.

\section{An account of phenotypic evolution}

\begin{figure*}[htp]
\centering
\includegraphics[width=17 cm]{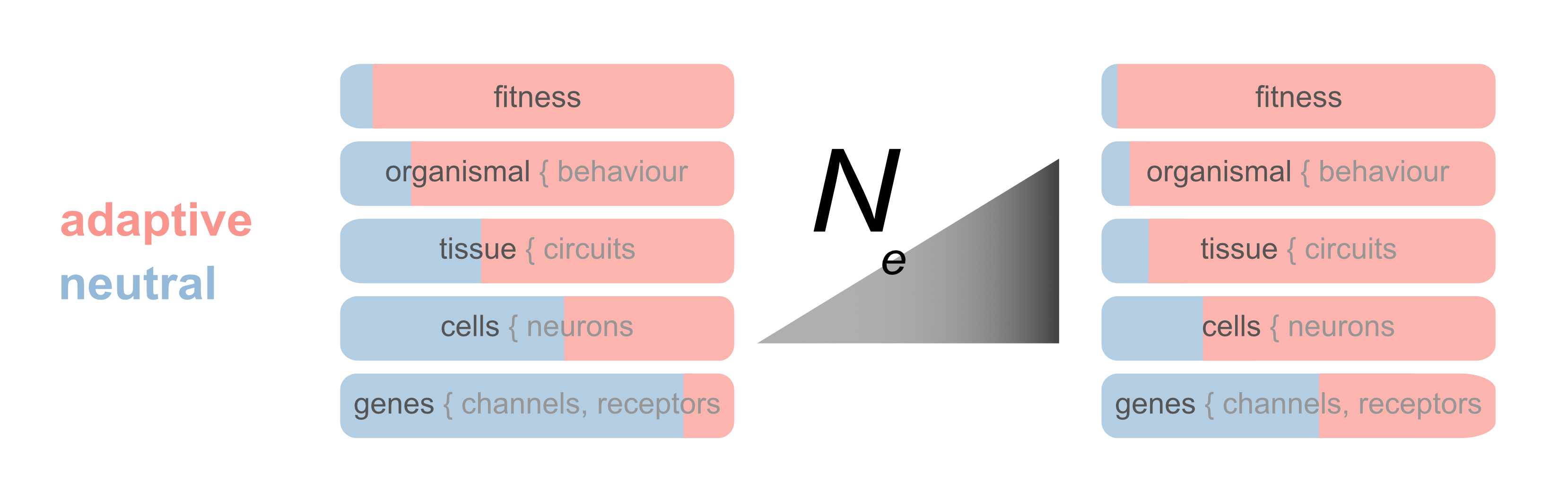}
\caption{\textbf{Higher phenotype traits in organisms with large effective population sizes are most susceptible to adaptive evolutionary change.} In contrast, lower phenotype traits in organisms with smaller effective population sizes are most likely the result of a drift-dominant evolutionary process. Each rectangle above represents a phenotypic `layer', and each layer consists of a fraction of changes from adaptive forces (in pink), and a complementary fraction of changes from neutral forces (in blue). The fraction of adaptive change increases as the level of phenotype is higher. The fraction of adaptive change increases across all phenotype levels as the effective population size increases.}
\label{fig_framework}
\end{figure*}

Nervous system-possessing animals have, relatively speaking, the smallest $N_e$ values. These values do not, however, suggest that those species are impervious to natural selection. The previous section's survey of efficient neural codes argues, if anything, that natural selection is well at work for nervous system-possessing animals. Indeed, we remind the reader that a given $N_e$ value should only be considered small relative to some selection coefficient, $s$, for that species, and that quantifying $s$ for the heritable basis of a species' neural coding scheme is, like for many complex traits, incredibly challenging and unknown. Rather, the purpose of highlighting the small $N_e$ values of nervous system-possessing animals is to argue that some species are less, but not totally, immune to stochastic evolutionary processes. As we saw in Section \ref{Empirical support for predictions on biological coding inefficiency}, unicellular eukaryotes, despite having one of the highest $N_e$ values, still exhibit inefficiencies in their genetic coding schemes not seen in their higher $N_e$ counterparts, such as the retention of genomic non-coding elements. For this reason, we think it is not unreasonable to expect the occasional inefficiency from species with the lowest $N_e$ values, like those with nervous systems. In this section, we outline a framework that will provide us with intuitions for when we might expect a biological trait to be the function of a stochastic historical process. We realize that this framework may be broadly applicable for biological observations outside of the neural sciences, but we think its impact might be particularly useful for neuroscience, because of the immense success that an efficiency-based worldview has had for the field. 

\subsection{Phenotypes can be stratified into a hierarchy}

Under what contexts might a non-adaptive historical process, as opposed to an adaptive one, better explain a neurobiological observation? For example, why do so many neural codes---in particular, ones at the periphery---lend themselves so easily to an efficiency-based worldview? Shouldn't the small $N_e$ values of nervous system-possessing animals lead us to expect inefficiencies across the nervous system? 

We argue that small $N_e$ values should not result in our thinking that non-adaptive forces exert themselves uniformly across all levels and traits of the species in question. That is, if a species' $N_e$ is small, the contribution of drift will be quite different, depending on whether we are trying to explain something like neural firing patterns, or gene expression. In some ways, this view is quite distinct from existing accounts of evolution. Most evolutionary frameworks express themselves exclusively in genetic terms. For example, in the allele fixation case from Section \ref{Random forces can dominate evolution}, the prevalence of a biological trait in a population is described in terms of its underlying genetics---that is, in terms of allele \textbf{A}'s fixation. In reality, however, one or two genes rarely encode for a phenotype, and theorizing about gene fixation dynamics is not identical to theorizing about the fixation dynamics of phenotypes that those genes contribute to. Although genetic material is the unit of inheritance and ultimately encodes for all biological processes, an account of evolution that is about the root cause is not the same as an account of evolution that is about the encoded products of the genetics. The central dogma of molecular biology \cite{crick_1970} explains how genetic material encodes for protein products, but it does not explain how those proteins encode for organismal function. Theories about genetic information are not equivalent to theories about phenotypes ultimately derived from genetic information, and explanations about genes do not, by mere substitution, extend to their products. Even Motoo Kimura, the founder of neutral evolutionary theory, speculated that the relevance of neutral theory is likely minimal, with respect to the evolution of phenotypes \cite{kimura1983neutral, kimura2020my, Zhang_2018}.

We contend that an account of evolution that at least acknowledges the gap between the evolution of genotypes and the evolution of phenotypes will best inform intuitions about when to appeal to adaptive or non-adaptive evolutionary explanations. More specifically, we propose the divvying of phenotypes into constituent `layers' \cite{Ho_Ohya_Zhang_2017, mayr_1997, wideman_doolittle_2019, Zhang_2018}. We posit that doing so will explain seeming discrepancies such as the observed efficiencies in neural coding in small $N_e$ species. The key advancement that a hierarchy of `layered' phenotypes provides is that elements of a bottom layer can afford to drift and be `inefficient', provided that their effects do not impinge on the integrity of the layer directly above. In other words, intermediary layers `buffer' the effects that bottom layers have on the highest layers, such as animal fitness. In this way, small $N_e$ species may possess many neutral mutations and their genetics may be subject, relatively strongly, to drift, but higher level physiologies such as their neural coding schemes, are still subject to strong natural selection, and contribute towards some form of efficient neural code. In contrast, a `layer-free' conception of phenotypes wherein genes directly encode for organismal attributes, results in an unrealistic and onerous emphasis of the genetic level's impact on species fitness. Such a conception is not compatible with existing biological evidence, because alleles that are slightly deleterious with respect to biochemical and cell-biological levels can exist, without dramatically compromising fitness \cite{wideman_doolittle_2019}. 

Given that natural selection acts directly on phenotype, rather than genotype \cite{mayr_1997}, the hierarchical stratification of phenotype, with respect to the genetic level, means that the relative contribution of natural selection on animal evolution is, in fact, a function of two parameters (Figure \ref{fig_framework}). One is the effective population size, $N_e$, as detailed above. The other is the phenotypic level at which species variation is observed. Such variation can range from the lowest level, genetic material, to the highest levels, such as behaviour and physiology. Change occurring at a level to which selection is blind will be neutral, provided that it does not affect the higher level \cite{mayr_1997, wideman_doolittle_2019, Zhang_2018}. By this definition, the probability of a given change being neutral is greater, the lower the level of the change. 

\subsection{A hierarchical view on phenotypes can explain differing efficiencies across species and phenotypes}

\begin{figure*}[htp]
\centering
\includegraphics[width=15 cm]{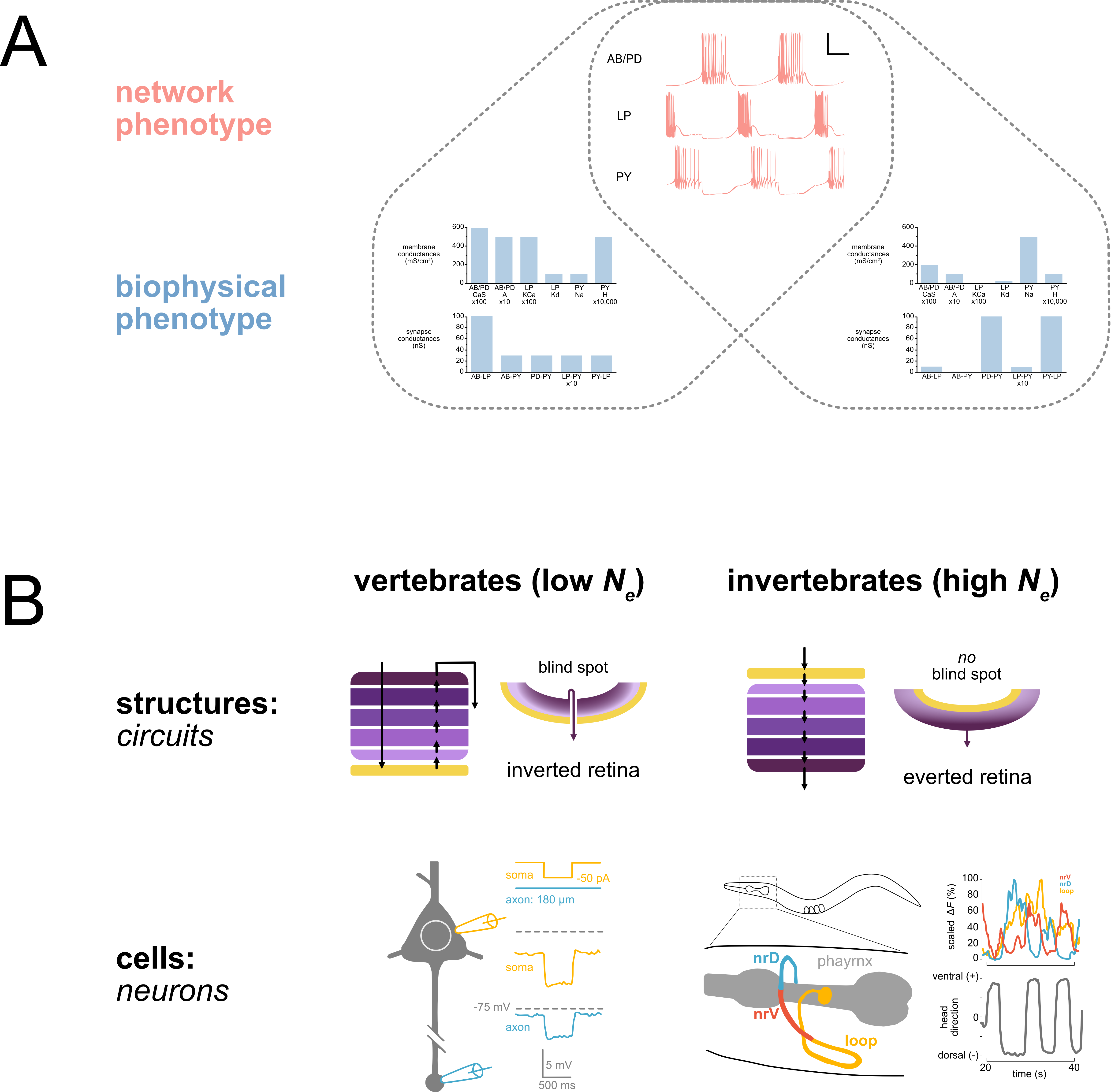}
\caption{\textbf{Differing contributions of adaptive and non-adaptive evolutionary forces can explain neurobiological phenomena.} A) For a given species, like the lobster, \textit{Homarus americanus}, multiple biophysical parameter combinations can realize the same higher level network phenotype. Lower phenotypic levels can afford to drift more than higher level phenotypes. This figure is a modified reproduction \cite{prinz_marder_2004}. B) For a given phenotype level, like a circuit, or a neuron, non-adaptive forces make stronger contributions to the evolution of species with smaller effective population sizes. Vertebrates possess inverted retina with blind spots and most vertebrate neurons are multipolar cells that perform single computations. Invertebrates possess everted retina without blind spots and most invertebrate neurons are unipolar and can perform multiple computations. For the retina schematics, arrows refer to the direction of information flow, and yellow refers to the photoreceptive cells. The non-retinal figures are modified reproductions \cite{hu_bean_2018, Hendricks_Ha_Maffey_Zhang_2012}.}
\label{fig_examples}
\end{figure*}

The above conception of phenotypes alters when we might appeal to adaptive or non-adaptive historical explanations, when thinking about the origins of neural codes. Despite the relatively small $N_e$ of nervous system-possessing animals, the efficiency of neural codes can be understood because of their close proximity to the highest level of phenotype. The fact that most efficient neural coding evidence relates to sensory processing---in particular, peripheral sensory processing---as well as the control of naturalistic behaviours, is likely not a coincidence. Neural codes instantiated in these systems interact directly with the native environment, and by extension, with selective forces. Indeed, in the case of the optimal facet angles of \textit{Hymenoptera} ommatidia \cite{barlow_1952}, the efficient implementation is not necessarily neural, but has in common the involvement of an apparatus that routinely interacts with the environment. The relevant parameter that explains the evolution of efficient neural codes is the proximity to phenotype, rather than it being neural, \textit{per se}. For this reason, the proposed framework also applies more broadly to general phenotypic attributes, such as gross animal morphology, and not just to nervous systems. That being said, we argue that a balanced evolutionary framework that incorporates both adaptive and non-adaptive processes is of special relevance to neuroscience, because of the unusual success that an exclusively optimization-based worldview has had for the field.

In this section, we interpret neurobiological results under the above articulated framework. We attempt to demonstrate, via examples, that a knowledge of both non-adaptive and adaptive evolutionary processes better explains the possible origins of neurobiological observations, than an appeal to adaptive processes alone. 

\subsubsection{Two levels of the lobster stomatogastric ganglion}

A clear example of how neutral forces may be more dominant in lower biological levels, but less so for higher phenotypes, can be seen in the lobster stomatogastric ganglion. This ganglion consists of a small and defined network of neurons that fire in a precisely ordered and periodic manner, so that the foregut of the lobster can contract rhythmically, in a so called pyloric rhythm, and enable the lobster's digestion of food. The simulation of more than 20 million biophysical models of the lobster's pyloric rhythm reveals that 20\% of these models can multiply realize the same pyloric-like rhythms. Importantly, these 4 million models, despite realizing the same network activity, were non-similar. Each model comprised of disparate biophysical parameter combinations \cite{prinz_marder_2004}.

For the neuroscientist limited to only the powers of natural selection, explaining the origins of such a finding is not necessarily straightforward. An appeal to adaptive processes alone struggles to explain the wanton biophysical degeneracy found in this ganglion. A complete understanding of evolutionary processes, however, provides some added explanatory value. Similar to the bottom-most level of genetic material, biophysical circuit parameters can afford to drift, provided they accomplish the same higher level phenotype---in this case, the network dynamics---required for organismal function and species fitness (Figure \ref{fig_examples}A). This example demonstrates the importance of stratifying phenotypes in a neuroscientific context. Natural selection is increasingly blind to bottom layer changes, which are effectively neutral, and need not be selected for or against. Bottom level changes can afford to drift because their ability to alter essential higher level phenotypes and viability is limited, and the relatively small $N_e$ values of nervous system-possessing lobsters predicts this drift. An appeal to adaptive historical processes is a more palatable hypothesis when explaining the evolution of higher, rather than lower, phenotypes.

\subsubsection{Vertebrate and invertebrate neurons}

The next two examples will compare and attempt to explain the differing efficiencies between vertebrate and invertebrate attributes. Given that the former has a smaller $N_e$ than the latter, we argue that when explaining comparisons about the same level of phenotype for both groups, an appeal to non-adaptive forces is less inappropriate for vertebrates. 

Vertebrate neurons are typically multi-polar\footnote{We emphasize that they are \textit{typically} multi-polar. They are not always multi-polar.}. That is, they resemble the structures often seen in textbooks, where some post-synaptic dendritic processes converge onto a cell body, before becoming a pre-synaptic axon. In this way, for a given neuron, neural impulses, or information flow, must pass through the cell body, in order to exit and pass on to a post-synaptic cell \cite{Goaillard_2020, kandel2013principles}. Simultaneous voltage recordings of a single mammalian cortical neuron's cell body and axon show correlated signals of identical sign and similar shape, but with a reduced amplitude and a slightly delayed onset for the axonal signal \cite{hu_bean_2018}. That is, the cell body integrates then propagates a single signal, with some minor decay, to the axon (Figure \ref{fig_examples}B). In this sense, multi-polar neurons typical of vertebrates are limited to a single computation, whose dynamics can be represented by a single cell body recording. 

In contrast, invertebrate neurons are almost always unipolar, meaning that the cell possesses only a single neurite. In such a morphology, the cell body is not a point of information convergence. Rather, from the cell body's one neurite, the neurite branches in many complex patterns, such that some arbours are pre-synpatic, and others are post-synaptic. Information flow can pass through various branches independent of others, and in this way, multiple computations can be spatially multi-plexed onto a single neuron. Two studies demonstrate this idea. The first is anatomical. The examination of spiking local interneurons in the locust metathoracic ganglion reveals two fields of neuropilar branches that a single process links together. One of these fields possess numerous finte neurites with relatively uniform diameters, and is located in a ventral area where afferents from a hind leg hair terminate. The other field possesses sparser more varicose neurites, and is located in a more dorsal area, alongside the neurites of many leg muscle motor neurons. The majority of the dorsal field are output synapses, although they can also receive inputs, whereas the ventral field consists mostly of input synapses \cite{watson_burrows_1985}. These two fields, both of which belong to a single local interneuron, likely participate in two completely distinct and compartmentalized functions. 

Neural recording data in support of this hypothesis can be seen in the invertebrate \textit{Caenorhabditis  elegans}. Calcium imaging reveals spatially compartmentalized and multi-plexed computations within a single interneuron, RIA. The RIA neuron consists of three axonal segments: the loop, the nrV, and the nrD (Figure \ref{fig_examples}B). Of these segments, calcium transients in the nrV and the nrD correlate with ventral and dorsal bends of the animal head, respectively. Although the loop segment does not correlate with any obvious behaviours, its dynamics were distinct from those of the nrV and nrD domains. The RIA neuron's cell body did not exhibit calcium responses \cite{Hendricks_Ha_Maffey_Zhang_2012}. This multi-plexing of computations across space in single neurons is largely absent from the multi-polar neurons typical of smaller $N_e$ species, such as vertebrates. A consideration of both effective population size and phenotypic level informs a context-dependent worldview for appealing to both adaptive and non-adaptive evolutionary processes. 

\subsubsection{Vertebrate and invertebrate optics} 
In addition to differing morphologies and computations at the single-cell level, vertebrates and invertebrates possess distinct anatomies. These grosser differences again suggest a greater contribution from neutral forces for smaller $N_e$ vertebrates, than for larger $N_e$ invertebrates. 

Namely, when compared to the invertebrate eye, the vertebrate eye appears inefficient. An intuitive and optimal design would place the photo-sensors at the front of the eye, and then pass those signals to neural processing cells behind the photo-sensors. In this way, the neurons would not interfere with the incoming light, and contribute to tissue scattering. Instead, the vertebrate retina lies directly on the optical path, such that the light passes through the ganglion cells, then the inner synaptic layer, then the amacrine cells, then the bipolar cells, then the horizontal cells, then the outer synaptic layer, then the photoreceptor cell bodies, and then finally, the actual photoreceptive rods and cones \cite{sterling_laughlin_principles, land_nilsson_2012}. The retina then processes the received light back out the way it came, such that the ganglion cells perform the final computations, despite being the first retinal cells the light encounters. 

This inverted configuration not only scatters the incoming light, but also results in a literal blind spot \cite{land_nilsson_2012}. In order for the retinal information to leave the eye and enter the central nervous system, it must, at some point, route through all the lower cell layers, including the rods and cones. The spot where the neural information leaves the eye cannot receive stimulus information, and the animal is blind in this region (Figure \ref{fig_examples}B). The obviously more efficient retinal layout that would both preclude a blind spot and mitigate light scattering from tissues would arrange the retina in the everted order. That is, it would place the photoreceptive cells at the front, relative to the incoming light, and the neurons behind the preceding cell would always do the subsequent computation \cite{land_nilsson_2012}. In doing so, the proceeding computations will not interfere with the light path, and routing the neural information out of the eye will not require transecting the retina. 

There is little reason to think that such an everted structure is not possible in nature. In fact, invertebrate eyes possess exactly this configuration (Figure \ref{fig_examples}B). The apposition compound eyes of most diurnal insects and the superposition compound eyes of nocturnal insects and deep-water crustaceans position their photoreceptive cells closest to the incoming light, and route the neural processing behind the photosensors, so that the information enters the central nervous system with no blind spot \cite{land_nilsson_2012}. This comparison is somewhat incommensurable, however, because compound eyes operate quite differently from the cornea-lens mechanism of terrestrial animals. Rather, we should compare the vertebrate terrestrial eye against the one major terrestrial invertebrate group with a cornea-lens structure: spiders. The anterior median eyes, or primary eyes, of spiders are everted, and as a result, they are not subject to the light scattering and blind spots seen in its vertebrate counterparts \cite{land_nilsson_2012, norgaard_2008}. Moreover, several ancient invertebrate phyla such as polychaetes \cite{richter_2010}, and both the scyphomedusae and hydromedusae jellyfish \cite{garm_2010} possess everted retina. Given the relative small $N_e$ values of the more recent vertebrate species, we speculate that the inverted retina may have originated from ancient large $N_e$ animals via a series of stochastic drift-driven contributions.

Despite the clear evidence for an inefficient vertebrate retina, neurobiologists continue to investigate the structure in terms of a design problem. That is, they hypothesize some goal for the system, and argue that retinal properties are subservient to the goal. For example, vision researchers have argued that the inverted retinal configuration affords some modest degree of space-saving efficiency, so that more retinal cells can be compacted in a given volume \cite{kroger_2009}. This argument, however, fails to address whether such a modest advantage is worth the light scattering and blind spot that comes with the inverted design. Indeed, the functional relevance of packing in more retinal cells, in general, is unclear. Similar arguments have been made regarding the retinal layers. Given that the retina lies on the optical path, neuroscientists have argued that the retina has optimized itself to be as thin as possible, so that it minimally scatters the incoming light \cite{sterling_laughlin_principles}. We emphasize, however, that an everted retina would circumvent this problem, as well as the blind spot. A consideration for neutral evolutionary processes may address this overcommitment to an efficiency-based framework. That being said, we do not disagree that natural selection has contributed significantly to retinal structures, including those of vertebrates. Several compelling arguments exist for the vertebrate retina's signal processing and general computational abilities. We merely aim to identify contexts where an appeal to neutral evolutionary forces may provide needed explanatory power. The small $N_e$ of vertebrates, relative to invertebrates, for a given phenotypic level, is one. 

\subsubsection{Speculations on brainwide activity across species}
How might a framework that acknowledges non-adaptive processes be relevant for explaining the origins of recent results in neuroscience? In this brief example, we provide interpretations of contemporary data to demonstrate the ongoing relevance of neutral evolutionary forces. 

Recent technological innovations have enabled neuroscientists to record at an unprecedented scale and resolution, across species. These brainwide recordings identify highly redundant representations of spontaneous ongoing behaviours, across the entire brain. In particular, smaller $N_e$ vertebrates appear to contribute a larger fraction of neural activity for representing their behaviours, than their larger $N_e$ counterparts, such that even dedicated visual processing areas, like V1, are found representing ongoing motor actions \cite{stringer_harris_2019, kato_2015, aimon_2019, schaffer_2021, kaplan_zimmer_2020}. In contrast, for the invertebrate \textit{C. elegans}, the sensory neurons fail to represent ongoing behaviours, and in this sense, do not contribute to the dominant redundant representation \cite{kato_2015, kaplan_zimmer_2020}. Similarly, in flies, although a significant fraction of brainwide activity represents ongoing behaviours, residual activity is high-dimensional, and appears to represent non-behavioural information, such as sensory cues and internal states \cite{aimon_2019, schaffer_2021}. Neutral evolutionary forces may, in part, explain the origins of this greater redundancy reduction for large $N_e$ invertebrates over small $N_e$ vertebrates. In general, for all species, the central nervous system's limited interaction with the natural environment, relative to the peripheral processing systems, may explain the subjection to possible neutral forces, and the resulting neural redundancy and lack of a clear objective function. We emphasize, however, that such redundancy does not mean natural selection does not operate, whatsoever. Indeed, for many macroevolutionary processes, the outcomes of neutral events can be co-opted for more adaptive means, such that they facilitate functional dependencies and neofunctionalizations \cite{Lynch_2007, bruckner_2021, kebschull_luo_2020}. Speculating on possible systems-level functions and goals is still essential for neuroscience. Evolutionary explanations are more powerful, however, when one can appeal to both non-adaptive and adaptive processes. 

\section{Non-adaptive explanations and the origins of neural codes}

Evolution explains the origins of biological things. It provides a process-based explanation for how a biological observation came to be. It can explain the efficiencies of neural codes, but it can also explain relative inefficiencies seen across phyla. Doing so, however, requires neuroscientists to not limit their evolutionary explanations to adaptive processes, like natural selection (Figure \ref{fig_explanations}B). We began this paper with a historical and speculative diagnosis as to why neuroscientists limit their explanations. The success of an efficiency-based worldview in neuroscience invites teleological implementations of an adaptive evolutionary process. To address this epistemological shortcoming, we introduced in this paper the concept of neutral stochastic evolutionary forces. We explained why the concept was first articulated, and we illustrate its accepted success for explaining biological features outside of neuroscience. We showed that evolutionary theory predicts that neutral processes make some non-negligible contribution to the origins of neurobiological features, and that, in addition, these processes comprise a useful epistemological tool for explaining neurobiological phenomena that natural selection struggles with. 

More precisely, we articulated a framework for providing intuitions about the relative contributions of adaptive and non-adaptive evolutionary forces, when thinking about the origins of a neurobiological feature. We theorize that the effective population size, just like for all of evolutionary biology, is a key parameter for hypothesizing the relative efficacy of natural selection on neurobiological traits. In addition, to $N_e$ values, we contend that the `level' of phenotype under evolutionary question is also germane for hypothesizing the relative contributions of drift. Phenotypes are the objects \textit{for} selection \cite{mayr_1997, sober1993nature}, because they are what directly interact with the environment. Not all phenotypes interact with the environment to the same degree, however, and for this reason, lower levels that are more obfuscated from the environment's effects are less subject to selective pressures. These lower levels, including the actual units of inheritance---the objects \textit{of} selection \cite{mayr_1997, sober1993nature}---are increasingly free to drift, because they are of less consequence to the species phenotypes that actually interact with the world. Even if the exact nature of this framework is questionable, we contend that one of the overall goals of this paper still stands: neurobiological features, like all biological features, are subject to the full repertoire of evolutionary forces, including non-adaptive and stochastic processes. Such processes can provide satisfactory explanations, just like natural selection can, and in some cases, more so. 

Evolution is a complex process. Rarely does it operate exclusively in adaptive or non-adaptive terms \cite{Gould_Lewontin_1979}. For this reason, we can still rightfully appeal to natural selection when explaining the many efficiencies of neural coding, but we can also speculate on the contributions of stochastic forces when we observe an obvious relative inefficiency. Neuroscientists must still hypothesize cost functions and goals for neural systems, but need not do so without acknowledging the non-teleological evolutionary means by which those functions arose. A careful consideration for all evolutionary forces has the greatest explanatory power. Evolution explains the origins of neural codes, including the occasional inefficiency. 

\newpage
\section*{Acknowledgements}
I thank  Bingni W. Brunton, Markus Meister, Joseph Parker, Zitong (Jerry) Wang, Alex Farhang, Kelly Kadlec, Guruprasad Raghavan, and Tara Chari for critical discussions and/or careful readings of the evolving manuscript. I thank Bingni W. Brunton for invaluable conversations and BLT sandwiches. I thank Markus Meister for teaching Principles of Neural Science (Bi154) at the California Institute of Technology and for encouraging me to share the paper with a broader audience. I am extremely grateful to Joseph Parker for his invaluable support, trust, and encouragement. I am supported by the Natural Sciences and Engineering Research Council of Canada's Postgraduate Scholarship. 

\section*{Author contributions}
Conceptualization: HK. Writing: HK. Editing and review: HK. Visualization: HK. 

\section*{Competing interests}
The authors have no competing interests.

\section*{Code availability}
Code for Figure \ref{fig_neutralforces}A and the original \LaTeX \ text are available at https://github.com/hanhanhan-kim/neuroevo\_paper. 

\newpage
\bibliographystyle{plain} 



\end{document}